\documentclass[%
 reprint,
superscriptaddress,
showpacs,
 amsmath,amssymb,
 aps,
 prl
]{revtex4-1}
\usepackage[caption=false]{subfig}
\usepackage{graphicx}
\usepackage{dcolumn}
\usepackage{bm}
\usepackage{hyperref}
\usepackage{float}

\begin{document}


\title{Optical levitation of high purity nanodiamonds  in vacuum without heating}%

\author{A. C. Frangeskou}
\email{a.c.frangeskou@warwick.ac.uk}
\affiliation{Department of Physics, University of Warwick, Gibbet Hill Road, Coventry CV4 7AL, United Kingdom}
\author{A. T. M. A. Rahman}
\affiliation{Department of Physics, University of Warwick, Gibbet Hill Road, Coventry CV4 7AL, United Kingdom}
\affiliation{Department of Physics and Astronomy, University College London, Gower Street, London WC1E 6BT, United Kingdom}
\author{L. Gines}
\author{S. Mandal}
\author{O. A. Williams}
\affiliation{School of Physics and Astronomy, Cardiff University, The Parade, Cardiff CF24 3AA, United Kingdom}
\author{P. F. Barker}
\affiliation{Department of Physics and Astronomy, University College London, Gower Street, London WC1E 6BT, United Kingdom}
\author{G. W. Morley}
\email{gavin.morley@warwick.ac.uk}
\affiliation{Department of Physics, University of Warwick, Gibbet Hill Road, Coventry CV4 7AL, United Kingdom}


\begin{abstract}
Levitated nanodiamonds containing nitrogen vacancy centres in high vacuum are a potential test bed for numerous phenomena in fundamental physics. However, experiments so far have been limited to low vacuum due to heating arising from optical absorption of the trapping laser. We show that milling pure diamond creates nanodiamonds that do not heat up as the optical intensity is raised above 700 GW/m$^2$ below 5 mbar of pressure. This advance now means that the level of attainable vacuum for nanodiamonds in optical dipole traps is no longer temperature limited.
\end{abstract}

\maketitle
Optically levitated nanodiamonds containing nitrogen vacancy (NV$^-$) centres have been proposed as probes of quantum gravity \cite{albrecht2014}, mesoscopic wavefunction collapse \cite{Scala2013,Yin2013,Wan2015,Wan2015b}, phonon mediated spin coupling \cite{albrecht2013}, and the direct detection of dark matter \cite{Riedel2013}. The NV$^-$ centre is a point defect in diamond that has a single electron spin which has long coherence times at room temperature and can be both polarized and read out optically \cite{Wrachtrup2006,Doherty2013}. Nanodiamonds containing NV$^-$ centres have been trapped at atmospheric pressure using ion traps \cite{Kuhlicke2014,Delord2016}, and a magneto-gravitational trap has allowed nanodiamond clusters to be held below $1\times 10^{-2}$ mbar \cite{Hsu2016a}. However, this design requires permanent magnets for the trapping, which is incompatible with the trap-and-release experiments that reach large distance spatial superpositions of the centre-of-mass as required for some of the fundamental physics experiments mentioned above \cite{albrecht2014,Wan2015b,Riedelb}. Progress with nanodiamonds levitated in optical dipole traps, where trap-and-release is possible,  includes the detection of NV$^-$ fluorescence \cite{Neukirch2013}, optically detected magnetic resonance  \cite{Neukirch2015,Hoang2015a}, and the observation of rotational vibration exceeding 1 MHz \cite{Hoang2016}. 

A key requirement of the aforementioned proposals is that the nanodiamonds are levitated in high vacuum to prevent motional decoherence arising from gas collisions. However, nanodiamond has been reported to heat up and eventually burn or graphitise below 10-40 mbar due to the absorption of trapping light by defects and impurities prevalent in the commercially available nanodiamonds used thus far \cite{Neukirch2015,Hoang2015a,Rahman2015}. Furthermore, heating has been shown to be detrimental to the fluorescence intensity of the NV$^-$ centre, which is necessary for the optical read out  of the spin state \cite{Toyli2012}. With non-levitated nanodiamonds, reducing the electron spin concentration has been shown to drastically improve spin coherence times of NV$^-$ centres in nanodiamond \cite{Knowles2014,Trusheim2013}. Since the source of absorption of near-infrared light by diamond is known to be extrinsic \cite{Dyer1965,Bennett2014,Webster2015}, improving the purity of the material used in optomechanics experiments involving nanodiamond to avoid heating and improve spin coherence is of vital importance to making the proposals in \cite{albrecht2014,albrecht2013,Scala2013,Yin2013,Wan2015,Wan2015b,Riedel2013} viable. We also note that heating is a problem more generally in optical trapping and not unique to nanodiamond \cite{Peterman2003,Millen2014}.

Here we report on levitated nanodiamonds milled from pure low nitrogen chemical vapour deposition (CVD) grown bulk diamond, and whilst methods such as reactive ion etching of diamond to form nano-pillars are known to produce superior quality nanodiamonds compared to milling ~\cite{Trusheim2013}, milling is the only technique we are aware of that produces large enough quantities of nanodiamonds for the common nanoparticle injection methods employed in optomechanics. A previous study had shown that commercial nanodiamonds (Ad\'{a}mas Nanotechnologies) can reach temperatures in excess of 800 K at 20 mbar, enough to burn or graphitise the nanodiamond \cite{Rahman2015}. This study had suggested that the source of heating was absorption of the trapping light by amorphous carbon on the surface of the nanodiamond and nitrogen defects in the diamond. Our milled CVD nanodiamonds remain at room temperature at lower pressures than the pressures attainable with commercially available material. These results confirm that nitrogen impurities are the dominant source of unwanted absorption and heating in commercially available nanodiamonds. We observe nanodiamonds to be suddenly ejected from the trap below 4 mbar (typically at $\sim$ 1 mbar), which we attribute to previously observed trap instabilities at intermediate vacuum that can be overcome with damping of the centre-of-mass motion \cite{Ranjit2015,Mestres2015,Gieseler2012}.


An optical dipole trap (optical tweezers) is formed by a focussed laser beam. A sub-wavelength sized dielectric bead satisfies the Rayleigh scattering criterion and can be approximated as a point dipole, and in the limit of small oscillations \cite{Gieseler2013} and the paraxial approximation \cite{Novotny2012}, the trap potential is harmonic with a spring constant
\begin{equation}
k_{trap} = 4\pi^3\frac{\alpha P}{c\varepsilon_0}\frac{\text{(N.A.)}^4}{\lambda^4}, 
\label{spring}
\end{equation}
where $\alpha$ is the polarisability of the bead, $P$ is the optical power, N.A. is the numerical aperture of the trapping lens, $c$ is the speed of light, $\varepsilon_0$ is the vacuum permittivity, and $\lambda$ is the trapping laser wavelength \cite{Gieseler2012}. Whilst the trap forms a harmonic potential, collisions with the surrounding gas induce Brownian dynamics \cite{Millen2014} and the motion of the bead is therefore governed by 
\begin{equation}
m\ddot x(t) + m\Gamma_0 \dot x(t) + m\omega_0^2x(t) = f_B(t),
\label{eqm}	
\end{equation}
where $x(t)$ is the time dependent position along the $x$ axis (transverse to the optical axis), $m$ is the mass, $\Gamma_0$ is the damping rate, $\omega_0 = \sqrt{k_{trap}/m}$, and $f_B(t)$ is a Gaussian random force with $\langle f_B(t) \rangle = 0$ and $\langle f_B(t)f_B(t-t') \rangle = 2m\Gamma_0k_{B}T_{cm}\delta(t-t')$, where $T_{cm}$ is the centre-of-mass temperature. Similar equations may be written for the $y$ (transverse) and  $z$ (along the optical axis) directions. It can then be shown that the  power spectral density is
\begin{equation}
S_x(\omega) = \frac{2k_BT_{cm}}{m}\frac{\Gamma_0}{(\omega^2 - \omega_0^2)^2 + \omega^2\Gamma_0^2}.
\label{psd}	
\end{equation}
We fit the experimental data with equation \ref{psd}, from which we may extract the beads centre-of-mass temperature, damping rate, size, and mechanical frequency. 

The centre-of-mass and internal temperatures are linked by $T_{cm} = (T_{imp}\Gamma_{imp}+T_{em}\Gamma_{em})/(\Gamma_{imp}+\Gamma_{em})$, where the subscripts $imp$ and $em$ denote the temperature and damping coefficients of impinging and emerging gas molecules, respectively. Gas molecules thermalise with the bead, and the temperature of the bead is then $\alpha_gT=T_{em}$, where $0\leq\alpha_g\leq1$ is the thermal accommodation coefficient which determines the degree of thermalisation \cite{Millen2014}. By measuring the centre-of-mass temperature of levitated Rayleigh beads as a function of trapping power, $P$, one may deduce whether the bead is heating, cooling, or at room temperature. We show that the centre-of-mass temperature of the majority of nanodiamonds investigated in this study showed no dependence on power, and were therefore still at room temperature at $\sim$4 mbar.


Approximately 150 mg of single crystal CVD bulk diamonds (Element Six 145-500-0274-01) were converted into nanodiamonds using silicon nitride ball milling. The nanodiamonds were cleaned with phosphoric acid at 180$^\circ$C and sodium hydroxide at 150$^\circ$C to remove the milling material, followed by a 5 hour 600$^\circ$C air anneal. Raman spectroscopy revealed no detectable contamination of silicon nitride (supplementary information). The concentration of single substitutional nitrogen defects (N$_\text{s}^0$) in 20 bulk samples of the same material were measured to vary from 95 ppb to 162 ppb using quantitative electron paramagnetic resonance \cite{EatonG.R.2010} (supplementary information). This signifies an increase in purity of approximately three orders of magnitude compared to the 150 ppm high-pressure high-temperature (HPHT) synthesised starting material used to make the nanodiamonds used for previous work \cite{Kuhlicke2014,Neukirch2013, Neukirch2015,Hoang2015a,Rahman2015,Hoang2016,Hsu2016a}.
 
\begin{figure} [t]
\includegraphics[width=0.48\textwidth]{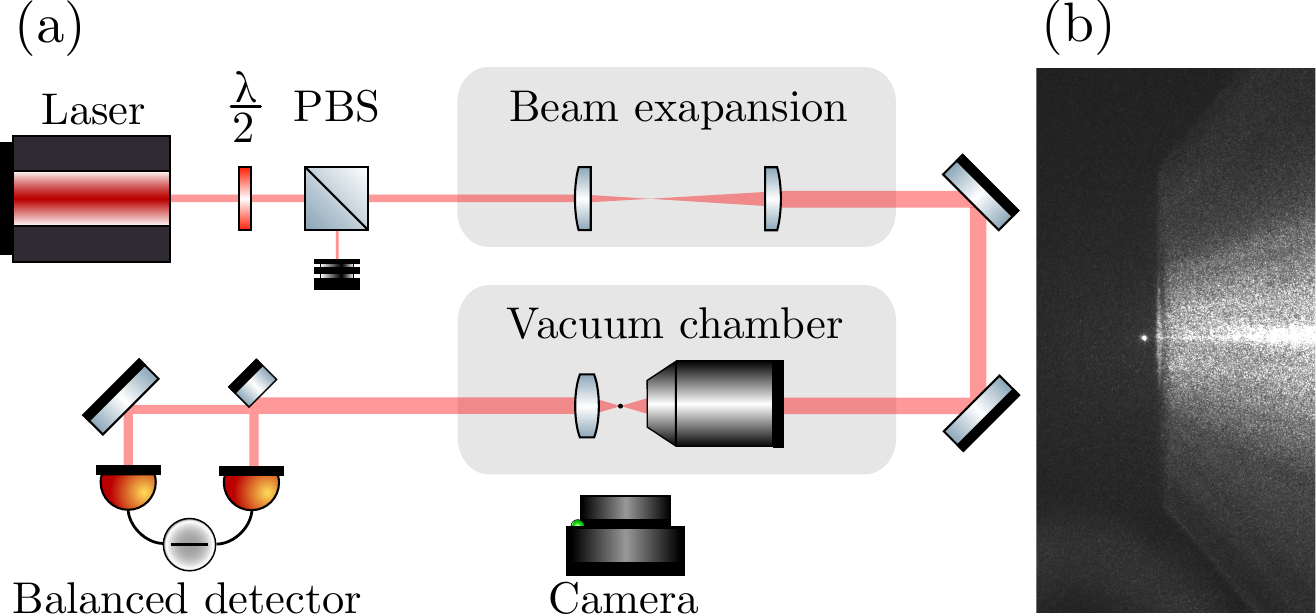}
\caption{\label{fig:schematic} (a) Schematic of the experimental setup. A half-wave plate ($\lambda/2$) and polarising beamsplitter (PBS) control the power of the laser. A microscope objective housed inside a vacuum chamber focusses the beam to form the dipole trap. Un-scattered and scattered light from trapped nanodiamonds is collimated by an aspheric lens and sent to a balanced detector using one half-mirror and one full mirror. A CMOS camera monitors scattering intensity from above the vacuum chamber. (b) Photograph of a levitated nanodiamond (white spot, left) trapped at the focus of the objective (right).}
\end{figure}

The optical dipole trap shown in Fig. \ref{fig:schematic} (a) was formed by focussing a single longitudinal mode 1064 nm Nd:YAG laser (Elforlight I4-700) with a microscope objective (Nikon N.A. 0.95) housed inside a vacuum chamber. The power of the laser was controlled with a half-wave plate and polarising beam-splitter, which transmits horizontally polarised light to the objective lens. Un-scattered and scattered light from the trapped nanodiamond was collimated by an aspheric lens and sent to an InGaAs balanced detector which monitors the $x$ motion in an interferometric scheme described in \cite{Gieseler2012}. 

\begin{figure}[!t]
\includegraphics[width=0.48\textwidth]{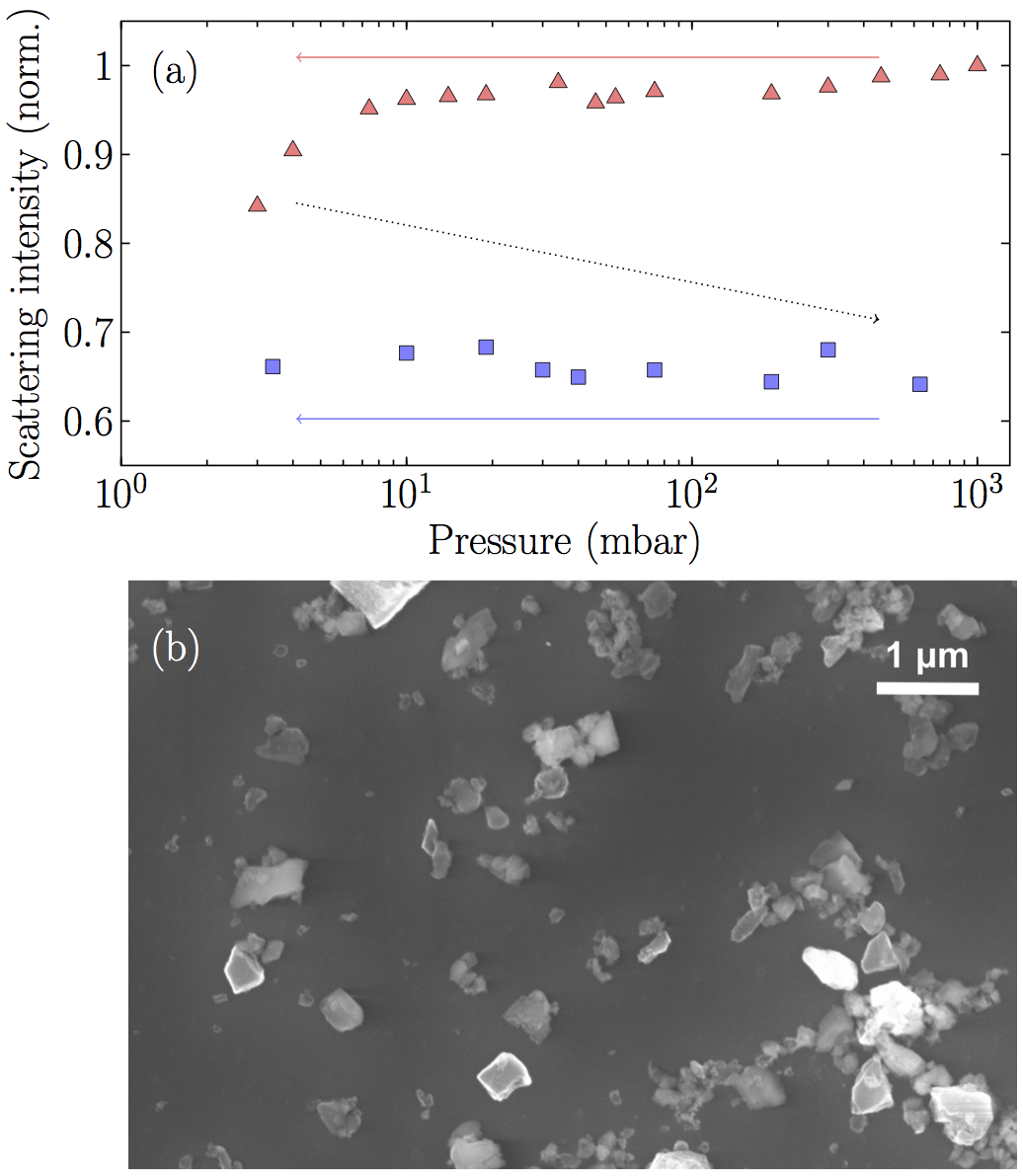}
\caption{\label{fig:nanodiamond}(a) Scattering intensity of a single nanodiamond as pressure is reduced from atmospheric to 4 mbar (red triangles). The chamber is then leaked back to atmospheric pressure and evacuated for a second time (blue squares). Minimum trapping power is used on the second evacuation, with the scattering intensity recorded after measurements to verify that the size of the nanodiamond has not changed during power variation. (b) SEM image of the nanodiamonds after ultra-sonication.}
\end{figure}

The nanodiamonds were suspended in pure methanol and sonicated prior to trapping (SEM image in Fig. ~\ref{fig:nanodiamond} (b)). Nanodiamonds were trapped by dispersing them into the vacuum chamber at atmospheric pressure using a nebuliser. Constancy of mass is a requirement of the power spectral density analysis, therefore nanodiamonds were first taken to  $\sim$3-4 mbar using the maximum available trapping power to remove surface contaminants \cite{Rahman2015}, and then brought back to atmospheric pressure after a minimum of one hour at vacuum. Scattered light from the nanodiamond was monitored with a CMOS camera above the vacuum chamber to ensure the size remained constant across all centre-of-mass measurements. Figure \ref{fig:nanodiamond} (a) shows that the scattering intensity falls by 10-20$\%$ after the initial evacuation due to the removal of surface contaminants, and then remain constant on the second evacuation when measurements were made \cite{Rahman2015}.

\begin{figure}[t]
\includegraphics[width=0.48\textwidth]{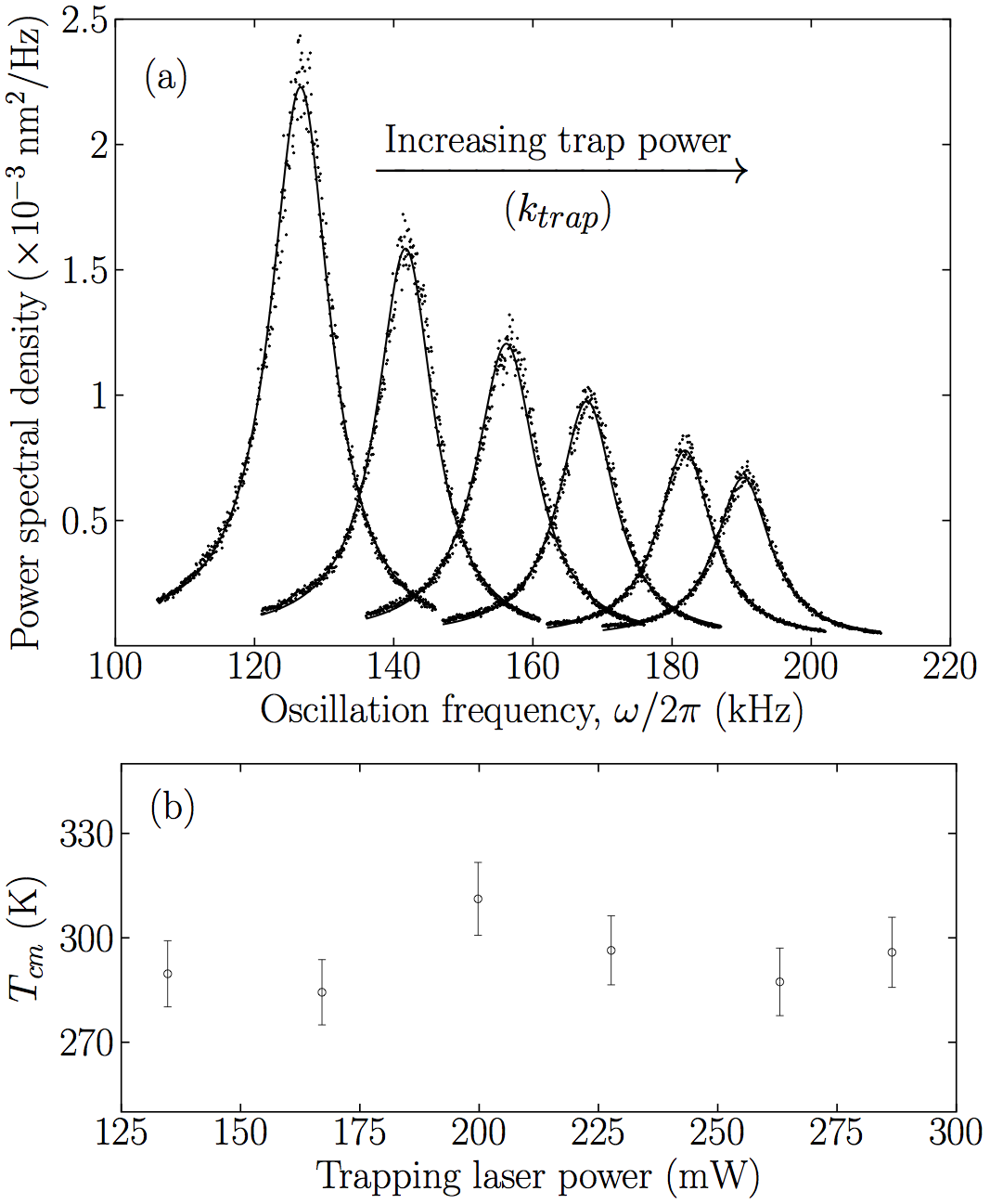}
\caption{\label{fits}(a) Experimentally measured power spectral densities (dots) of a single trapped nanodiamond at 4 mbar as trapping power is increased from 130 to 290 mW (left to right), fit with equation ~\ref{psd} (solid lines). As expected from equation \ref{spring}, the trap frequency increases as $\omega_0\sim\sqrt{P}$. From the fit, we determine the radius of the nanodiamond to be $r\approx 18$nm (equivalent sphere). (b) Corresponding centre-of-mass temperatures of the same nanodiamond, showing no dependence on the trapping laser power.}
\end{figure}

The live position signal from the balanced detector was recorded with a high resolution PC oscilloscope. After normalising by the power, the data is Fourier transformed to reveal the power spectral density and fit with $S_x(\omega)$ as shown in Fig. \ref{fits} (a). From the fit,  $A = 2C^2k_BT_{cm}/m$ was extracted for different powers at a fixed pressure of 4 mbar, where $C$ is a calibration constant for converting from the detector signal in volts to meters. By demonstrating that $A$ does not depend on power for a fixed mass, we confirm that the nanodiamonds remained at room temperature as the trapping power was more than doubled up to approximately 300 mW, corresponding to an intensity at the focus of $\sim$750 GW/m$^2$. The centre-of-mass temperature was measured as a function of power rather than pressure to take advantage of the increased signal-to-noise at low pressure. The centre-of-mass temperature was inferred from $A$ by measuring $A$ at a higher pressure $p_1$ where the nanodiamond is confirmed to be at room temperature, and then $T_{cm} = T_0(A_{p_1}/A_{p_2})$, where $T_0=298$ K, and $A_{p_1}$ and $A_{p_2}$ are the values of $A$ at low pressure $p_2$ and high pressure respectively.  

Representative data of the nanodiamonds studied in this article are presented in Fig. \ref{fits}. We demonstrate that the centre-of-mass temperature does not depend on the trapping laser power. This shows that by absorbing less trapping light, the nanodiamonds are able to dissipate their excess heat. Furthermore, we were able to keep most nanodiamonds trapped at $\sim$ 2.5-4 mbar for over a week with the maximum available trapping power. 

The upper-bound absorption coefficient at 1064 nm of the parent material used to make our nanodiamonds is 0.03 cm$^{-1}$ \cite{Bennett2014,Bennett2016b}, corresponding to an absorption index (i.e. the imaginary part of the complex refractive index) of $2.5 \times 10^{-7}$ i.e., comparable to the known absorption index of pure silica \cite{Kitamura2007}, which has already been optically trapped at high vacuum in numerous studies \cite{Li2011,Gieseler2012,Ranjit2015,Mestres2015,Jain2016,Vovrosh2016}. Therefore, if the motion of the nanodiamond can be damped through the period of intermediate vacuum, it may be possible to take the nanodiamonds to high vacuum where proposals \cite{albrecht2014,albrecht2013,Scala2013,Yin2013,Wan2015,Wan2015b,Riedel2013}  could be realised. We also note that single crystal CVD diamond with over 2 orders of magnitude greater purity than the nanodiamonds used here is commercially available.

\begin{figure}[t]
\includegraphics[width=0.48\textwidth]{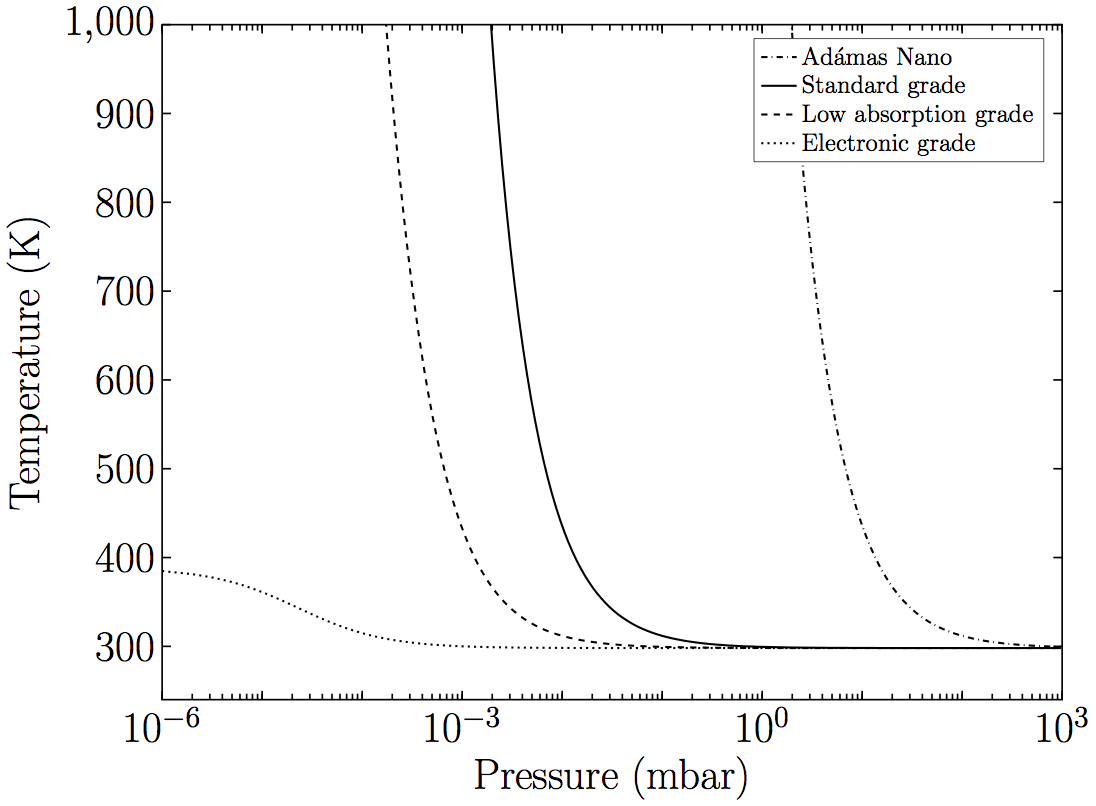}
\caption{\label{sim} Modelled upper-bounds of temperature as a function of pressure for $r=20$ nm nanodiamonds and trapping laser intensity of $I=60$ GW/m$^2$ \cite{Hoang2015a}. The absorption coefficient of the Ad\'{a}mas Nanotechnologies nanodiamonds (dot-dash line) used in \cite{Kuhlicke2014,Neukirch2013, Neukirch2015,Hoang2015a,Rahman2015,Hoang2016,Hsu2016a} has been set to $30$ cm$^{-1}$ based on 150 ppm N$_\text{s}^0$. Standard grade (solid line) corresponds to the bulk diamond grade used to make the nanodiamonds used in this paper, with an upper-bound absorption coefficient $0.03$ cm$^{-1}$. Low absorption grade (dashed line) has $0.003$ cm$^{-1}$ \cite{Bennett2014,Webster2015}. Electronic grade (dotted line) has a predicted absorption coefficient of $4.5\times 10^{-5}$ cm$^{-1}$.}
\end{figure}

In order to predict an upper-bound to the pressure our nanodiamonds can reach, we utilise a thermodynamic model based on two heat dissipation mechanisms: gas cooling and black-body radiation. Equation \ref{sim} models a sub-wavelength sphere with an absorption cross-section related to the complex permittivity of the sphere, a cooling rate due to gas molecule collisions, and both the absorption and dissipation of black-body radiation (see supplementary information). The excess heat remaining in the sphere can be expressed as \cite{Bohren1983,Chang2010}
\begin{eqnarray}
C_VV(T-T_0) = \underbrace{3IkV\left (\text{Im}\frac{\epsilon-1}{\epsilon+2}\right)}_{\text{Absorption}} \\*  \nonumber-\underbrace{6\alpha_g \pi r^2\bar{v}N_0\frac{p}{p_0}k_B\left(T-T_0\right)}_{\text{Gas}} \\*  \nonumber - \underbrace{\frac{72\zeta(5)V}{\pi^2c^3\hbar^4}\left(\text{Im}\frac{\epsilon_{bb}-1}{\epsilon_{bb}+2}\right)k_B^5T^5}_{\text{Black-body}},
\end{eqnarray}
where $C_V$ is the volumetric heat capacity, $I$ is the trapping laser intensity, $k=2\pi/\lambda$, $V$ is the nanoparticle volume, $\epsilon$ is the complex permittivity, $\alpha_g$ is the thermal accommodation coefficient, $r$ is the nanoparticle radius, $\bar{v}$ is the mean gas velocity, $N_0$ is the number of gas molecules at atmospheric pressure, $p$ is the gas pressure, $p_0$ is atmospheric pressure, $\zeta(5)$ is the Reimann zeta function, and $\hbar$ is Planck's constant divided by $2\pi$. It is assumed that $\epsilon_{bb}\approx\epsilon$ \cite{Chang2010}.

We model the steady state temperatures as a function of pressure of 20 nm radius nanodiamonds using upper-bounds of the absorption coefficients measured in \cite{Bennett2014,Webster2015} by laser calorimetry. Assuming a linear relationship between defect concentration and absorption coefficient \cite{Morelli1993}, we may also model diamonds for which the absorption coefficient is below the detection limit of laser calorimetry (0.001 cm$^{-1}$ for a 1 mm thick sample \cite{Webster2015}). Figure \ref{sim} shows the predicted temperature as a function of pressure for nanodiamonds of various absorption coefficients.

No significant heating is expected from 1 to 0.001 bar, as confirmed by our experimental results. At atmospheric pressure, nanodiamond graphitises at $\sim$ 940 - 1070 K \cite{Jian1999,Xu2002}. Taking 940 K as a cut-off, we estimate an attainable pressure of at least ${10^{-3}}$ mbar for the nanodiamonds used in this article, or ${10^{-4}}$ mbar for 20-30 ppb N$_\text{s}^0$ low absorption grade material. To reach the pressures required for proposals \cite{albrecht2014,albrecht2013,Scala2013,Yin2013,Wan2015,Wan2015b,Riedel2013}, electronic grade diamond containing less than 5 ppb N$_\text{s}^0$ may be required, which is available commercially. 

The actual attainable pressures are likely to be lower than this, as we have used the upper-bounds of the absorption coefficients. It is also likely that the absorption will differ at the nanoscale compared to the bulk. For an average particle radius of 25 nm, as is the case in this study, there are likely to be only 1 or 2 N$_\text{s}^0$ defects per nanodiamond if the concentration of N$_\text{s}^0$ is 100-160 ppb. In this regime, the surface is probably a more significant source of absorption than the bulk. We also note that high vacuum is commonly used in high-temperature annealing of diamond to prevent the onset of graphitisation, which may further extend the level of attainable vacuum \cite{Chu2014}. Since nanodiamonds are non-spherical (Fig. \ref{fig:nanodiamond} (b)), they have a higher surface area-to-volume ratio, which assists gas cooling. It should also be noted that a previous study has shown that the spin coherence lifetime of the NV$^-$ centre ($T_2^*$) can be over 1 $\mu$s even for temperatures above 600 K \cite{Toyli2012}. 

Although we were unable to measure the temperature at the loss pressure, the simulation shows that heating would be limited to 10 K above room temperature at 1 mbar at most (using the highest intensities and bead radius), which would suggest a different loss mechanism to the one proposed in \cite{Ranjit2015} involving radiometric forces arising from temperature gradients across 3 $\mu$m silica spheres. Given diamonds large thermal conductivity, temperature gradients are also unlikely to be present. Rather, the smaller damping coefficient at lower pressure, combined with nonconservative scattering forces \cite{Roichman2008,Wu2009}, also proposed in \cite{Ranjit2015}, and shot-noise from the gas, are the more probable culprits in this trapping regime. 

In conclusion, we have milled pure CVD diamonds into nanodiamonds and measured their resulting centre-of-mass motion in an optical dipole trap, from which we infer that the nanodiamonds do not heat up at a few mbar for the first time. We attribute the previously observed heating in commercial nanodiamonds predominantly to nitrogen defects within the diamond. We have therefore demonstrated a route to levitating nanodiamonds in high vacuum and have set upper-bounds on the pressure they can sustain by modelling the temperature of various types of nanodiamonds as a function of pressure. This advance makes experiments in dark matter detection, quantum gravity, phonon mediated coupling of electron spins, and matter-wave interferometry viable.

We acknowledge the EPSRC grants EP/J014664/1 and EP/J500045/1, and the European Union Seventh Framework Programme (FP7/2007-2013) under grant agreement no. 618078. G.W.M. is supported by the Royal Society. We are grateful to James Millen for comments that improved the manuscript, and to Ben Breeze for assistance with EPR.

\bibliography{library2}
\end{document}